\documentclass[10pt,a4paper]{article}
\usepackage[english]{babel}
\begin{document}
\begin{center}
{\bf \Large
Firm Projects and Social Behaviour of Investors
}
\end{center}

\begin{center}
Jana HUDAKOVA
\end{center}

\begin{center}
Matematical - Economical Consultancy \\
Stierova 23 \\
SK-040 11 Kosice, Slovak Republic

\end{center}

\begin{center}
Ondrej HUDAK
\end{center}

\begin{center}
Faculty of Finance, Matej Bel University, \\ Cesta na amfiteater 1, SK-974 15 Banska Bystrica, Slovak Republic,\\
e-mail: ondrej.hudak@umb.sk
\end{center}

\newpage
\section*{Abstract}

An investor is estimating net present value of a firm project and performs risk analysis. Usually it is created portfolio hierarchies and make comparison of variants of project based on these hierarchies. Then one finds that portfolio which corresponds to the particular needs of individual groups within the firm. We have formulated a new type of NPV analysis based on the fact that normal distribution of NPV is observed for some projects in some industries. The expected risk of the project is given by variance, in which there is $ \sigma_{n} $ the standard deviation of the year n cash flow, $ \sigma_{I} $ the standard deviation of the investment I in the time zero, $ \rho_{I,n} $ the correlation coefficient of the year n cash flow deviation from the average and of the investment I at time zero deviation from the mean investment at time zero, $ \rho_{n,n´} $ the correlation coefficient of the year $ n $ cash flow deviation from the average and of the year $ n´$ cash flow deviation from the average. The aim function of the investor into the project was found. The investor is characterized by the constant A. The larger constant A the larger preference is given to the project NPV and the larger acceptable risk of the project, and vice versa. We have found that there are contributions in which we have the aim-function-like contribution to the aim function, which is discounted and in which the risk of the n-th year risk is discounted in the second order. Further there is aim-function-like contribution to the aim-function which comes from the initial investment I and its risk $ \sigma^{2}_{I} $.

\newpage
\section{Introduction}

A social group of investors investing in a firm project and of buyers buying products of the firm project is characterized as any general group \cite{1} - \cite{5}.  We assume that individual behaviour
of investors and buyers is influenced by the need to associate with other investors and buyers (agents) to obtain the approval of other agents in the group of agents which is characterized by large nonrational
and emotional element in decisions. Making decisions an agent equates own needs with those
of the other agents from the group. Agents from the group may interact. The interaction of investors and buyers consists of the exchange of information and it costs some money. The exchanged information is assumed to be well defined, and we assume that agents interact in such a way that they give reference
to the origin of the information if asked by other agents.
The information is thus private \cite{6}.

\section{Net Present Value of the Project}

An investor is estimating NPV (net present value) of a firm project. Then he/she performs risk analysis. In \cite{SAP} it is described for example how software SAP for Banking controls risk on the market. Static and dynamic risk control of market risk is focused either on the NPV calculations or on periodic results. The software enables to manage the risk for interest rate changes, currency fluctuations, market volatility and other quantities. Analysing of NPV and Value at Risk (VaR) the software measures and controls market risk using different simulation scenarios. Quick recognition of any profit risk and profit-and-loss calculations is based on combining gap analysis and strategy management possibilities included in this software. It is possible to obtain the balance sheet view of risk. Gap analysis enables to control the liquidity risk. Static analysis of market risk for the firm (bank) is based on existing portfolios or exposures. The software enable to calculate NPV for different bank products, then using scenarios for development of yield curves, volatility or exchange rate simulations, it is possible to perform VaR analyses (Monte Carlo f.e.), historic simulations, variance and covariance calculations. This scenarios may be tested on the historical data.
Then it is possible to create portfolio hierarchies and make comparison based on these hierarchies. Then one can use this analysis to find that portfolio which corresponds to the particular needs of individual groups within the firm (bank). Thus this software provides consistent management of the market and default risk, see \cite{SAP}.
As we can see the software enable to create various scenarios of development and to choose that product which corresponds to the needs of the group.

Recently \cite{NPVAR} published NPV-at-Risk method in infrastructure project investment evaluation. Authors note that an investment in privately financed infrastructure projects requires careful consideration which take into account high levels of financial, political, and market risks. A systematic classification of existing evaluation methods by the authors has shown that it is possible to develop a new method — the net-present-value-at-risk (NPV-at-risk) method by combining the weighted average cost of capital and dual risk-return method. The authors evaluate two hypothetical power projects and have shown that their method can provide a better decision for risk evaluation of, and investment in, privately financed infrastructure projects. For systematic classification of existing evaluation methods see in this paper.

There are evidences that distribution of Net Present Value (NPV) of the same projects in the same industry has normal distibution \cite{ND1} - \cite{ND21}
However there exists also evidence that this distribution may be in some cases different from the normal distribution, see the literature cited for discussion. 
Then there is an interesting problem: how the behaviour of distribution of NPV of a firm projects may be characterized for a given firm project in a given industry. Another interesting problem is how to describe NPV of a firm project taking into account the risk of the project using characteristics of the distribution of NPV of firm projects.

While the first problem is difficult, there is not enough data for various firm projects in various industries, the second problem may be solved for those firm project NPV of which is normally distributed. So in this paper we will assume that the distribution of NPV of firm projects is normal. It is however clear that this is not true for all possible projects, see references above for discussion this point.

One can ask now whether an opposite type of analysis to those described above does not correspond better to reality. We will not develop various scenarios and then choose the most convenient for the investor, we will assume that the group (firm) is characterized by a risk aversion characteristics, which is intrinsic to this group (firm) and then using an aim function we will perform the analysis of the project in which the NPV, its risk and an aversion of the investor are included. This type of analysis is used in fact for analysis of NPV for investment into shares and bonds on capital markets. In general an investment on a capital market may be considered also  as a firm project. It is known that on liquid developed markets in equilibrium there exists equivalence between investing into the shares financing a given firm project and investing directly into the firm project. 
In this paper we thus formulate such a type of analysis of NPV and risk of investment into project which is based on the aim function as in the case of shares in the Markowitz model \cite{7} - \cite{10}.

It is known that there exists normal distribution of NPV of projects in some industries, see in \cite{ND1} - \cite{ND21}. In some case this distribution seems not to be normal as we mentioned. We will consider the first case here in this paper only.

It is assumed that, for simplicity,
there exists an investment I /which is a negative value if it is an investment of the investor for which other sources of investment like grants, subsidiarisation, etc., are less then his own sources, and it is positive if it is an inverse situation/ in time zero, and that there is a cash flow $CF_{n} $ from the project every n-th year. The project life-time is assumed to be N years. Then the NPV of such an investment is

\begin{equation}
\label{0.1}
NPV = \sum_{n=1}^{n=N} \frac{CF_{n}}{(1+r)^{n}} + I
\end{equation}

The estimation of NPV in (\ref{0.1}) is usually such that it is assumed that cash flow $ CF_{n} $ in the year n has certainty to realize in the future with probability 1.  It is assumed that the investment I has certainty to realize in the near future /time zero/ with probability 1. This is a simplification, neverthless this assumption is used to be used.
This assumption is however in general not true. Investors then have to perform Risk and Sensitivity Analysis of the type described above, or similar with various scenarios, however assuming certainty to realize the investment and the cash flows with probability 1 in every possible case of scenario. We do not consider here exclusive cases of an investment and cash flows. Then investors consider variations of the investment and cash flows for the same case.
To consider which project is for a given investor the most profitable at a given expected risk, or which project is the less risky for an expected profit one should consider uncertainty in looking for such an optimal project. Uncertainty in cash flows and /in principle also in the investment in the time zero/ lead investors in fact to work with expected cash flows for the future periods, and with expected investment I in zero time. Expectations of investors should include not only expected cash flow for every year n and for the expected investment I in the zero-th year, but also expected risk of net present value $ \sigma^{2}_{NPV} $.
 Most of the projects has normal distribution of NPV as we mentioned above. So we will assume in the following in this paper that this distribution of NPV for projects /in a given industry/ is a normal distribution. Note that NPV of a project contains for every year cash flow which is usually positive /however it may be also negative/ and the initial investment is a negative return usually /however, as we already discussed it may be also a positive quantity/.

There exists a mean value of the NPV for projects of the same type in a given industry because we consider projects of the same type in one industry for simplicity. More complicated cases of different projects in the same industry, and of projects in different industries may be described generalizing our description assuming, that there is a liquid market with projects. Less liquid market of projects we do not consider here. Note that the mean value of the NPV, $ \overline{NPV} $ of the project is given by

\begin{equation}
\label{0.2}
\overline{NPV} = \sum_{n=1}^{n=N} \frac{\overline{CF}_{n}}{(1+r)^{n}} + \overline{I}
\end{equation}

\section{The Risk of the Net Present Value of the Project}

The risk of the NPV of the project is for normal distribution of NPV characterized by a variance of the NPV of the project. If we consider M time periods over which we calculate the variance of the project /we consider an industry in which there are investments in the project of the same type over longer time period - M time periods/. Then the variance is defined as

\begin{equation}
\label{0.3}
\sigma^{2} = \frac{1}{M-1} \sum_{i=1}^{m=M} (NPV_{i} - \overline{NPV} )^{2}
\end{equation}

Here $ NPV_{i} $ is the NPV of the project in time i. We assume the liquid market with projects, so we can assume that at every time i there is a possibility to invest in the project. We will average over projects which start in different time i. There is a rare case that the same project in the same industry starts exactly at the same time i. So we will not consider this possibility.

We can rewrite the difference in the definition of the variance

\begin{equation}
\label{0.4}
NPV_{i} - \overline{NPV} = \sum_{n=1}^{n=N} \frac{CF_{n}^{i} - \overline{CF}_{n}}
{(1+r)^{n}} + (I_{i} - \overline{I})
\end{equation}

where $ CF_{n}^{i} $ is the expected value of the cash flow in the year n and in the time i /time on the market with the project/, and where $ I_{i} $ is the expected investment into the project in time i /time on the market with the project/. Here $ \overline{I} $ is the mean value of the investment I into the project over the time on the market with the project.
The quantity $ CF_{n}^{i} $ is the value of the cash flow CF into the project as it is expected in the year n and i denotes that we consider the time on the market with the project.

The expected risk of the project is then, substituting from (\ref{0.4}) into (\ref{0.3}), written in the form

\begin{equation}
\label{0.5}
\sigma^{2} = \frac{1}{M-1} \sum_{i=1}^{m=M} ( \sum_{n=1}^{n=N}
[ \frac{(CF_{n}^{i} - \overline{CF}_{n})^{2}}{(1+r)^{2n}}
\end{equation}
\[ + (I_{i} - \overline{I})^{2} - 2 (I_{i} - \overline{I})
\frac{CF_{n}^{i} - \overline{CF}_{n}}{(1+r)^{n}} ] + \]
\[ 2 \sum_{n=1, n´>n}^{n=N. n´=N}  \frac{(CF_{n}^{i} - \overline{CF}_{n})(CF_{n´}^{i} - \overline{CF}_{n´})}{(1+r)^{n+n´}}  \]

The expected risk of the project may be written from (\ref{0.5}) in the form

\begin{equation}
\label{0.6}
\sigma^{2} =  \sum_{n=1}^{n=N}  \frac{{\sigma_{n}}^{2}}{(1+r)^{2n}} + \sigma^{2}_{I}
 - 2 \sum_{n=1}^{n=N}  \rho_{I,n} \sigma_{I} \sigma_{n}\frac{1}{(1+r)^{n}}
  + 2 \sum_{n=1, n´>n}^{n=N. n´=N}  \rho_{n,n} \sigma_{n} \sigma_{n´}
  \frac{1}{(1+r)^{n+n´}}
\end{equation}

where $ \sigma_{n} $ is the standard deviation of the year n cash flow, $ \sigma_{I} $ is the standard deviation of the investment I in the time zero, $ \rho_{I,n} $ is the correlation coefficient of the year n cash flow deviation from the average and of the investment I at time zero deviation from the mean investment at time zero, $ \rho_{n,n´} $ is the correlation coefficient of the year $ n $ cash flow deviation from the average and of the year $ n´ $ cash flow deviation from the average.

We can see from the (\ref{0.6}) that the risk of the cash flow and the zero time investment
is a sum  of the discounted risk of the risk of the year n cash flow, of the zero time investment risk, of the risk associated with the correlation of the zero time investment I and the year $ n $ cash flow and  of the risk associated with the correlation of the year $ n´ $ cash flow and the year $ n $ cash flow. The later two terms are discounted correspondingly. Thus the risk of the NPV of the project consists of discounted contributions of the cash flow risks and correlations. The larger the rate of return in discounting coefficient, the smaller contribution from the risk of the year n cash flow, from the year n cash flow deviation from the average correlation with the zero time investment I deviation from the average and
from the year n cash flow deviation from the average correlation with from the year $ n´$ cash flow deviation. The smaller the rate of return in discounting coefficient, the larger contribution from the risk of the year n cash flow, from the year n cash flow deviation from the average correlation with the zero time investment I deviation from the average and
from the year n cash flow deviation from the average correlation with from the year $ n´$ cash flow deviation.

\section{The Aim Function of the Investor into the Project}

The investor has an expected NPV of the project. To estimate this expected value we will use the historical values of NPV of the project in the industry. To describe the expected risk of the project we will use the historical value of the risk of the project. We assume thus that expected NPV and expected risk of the project do not differ too much from
recent historical data on projects with N years of their life. This assumption is good for projects for which N is larger than let us say two business cycles. This assumption is
good also for projects in which N is two-three months. The assumption is
expected not to be good for projects in which N is nearly one to three years because the business cycle leads to different behaviour of such projects in recent period with respect to the period in the next one to three years.

The investor is characterized by the constant A, which is nonnegative. The larger constant A the larger preference is given to the project NPV and the larger acceptable risk of the project.
The smaller constant A the smaller preference is given to the project NPV and the smaller acceptable risk of the project.

The aim function has the form

\begin{equation}
\label{0.7}
F = - A. NPV +  \sigma^{2}
\end{equation}

The aim function can be rewriten now into the form using
(\ref{0.1}), (\ref{0.5}) and (\ref{0.7}):

\begin{equation}
\label{0.8}
F = - A. (\sum_{n=1}^{n=N} \frac{CF_{n}}{(1+r)^{n}} + I)
 +  \sigma^{2}
\end{equation}

and

\begin{equation}
\label{0.9}
F =  \sum_{n=1}^{n=N} \frac{{1}}{(1+r)^{n}} ( - A. CF_{n} + \frac{{\sigma_{n}}^{2}}{(1+r)^{n}} ) + ( A.I + \sigma^{2}_{I})
 - 2 \sum_{n=1}^{n=N}  \rho_{I,n} \sigma_{I} \sigma_{n}\frac{1}{(1+r)^{n}}
  + 2 \sum_{n=1, n´>n}^{n=N. n´=N}  \rho_{n,n} \sigma_{n} \sigma_{n´}
  \frac{1}{(1+r)^{n+n´}}
\end{equation}

As we can see there are contributions in the sum which have the aim-function-like contribution form to the aim function, which is discounted and in which the risk of the n-th year risk is discounted in the second order. This is the first term in (\ref{0.9}). The second term 
in (\ref{0.9}) contains aim-function-like contribution to the aim-function which comes from the initial investment I and its risk $ \sigma^{2}_{I} $. However the investment is a negative cash flow (expected). The third term in (\ref{0.9}) contains the contribution of the cash flow risk (as standard deviation) in the n-th year and the investment risk (as standard deviation) and the correlation coefficient $ \rho_{I,n} $ between the returns of the investments I in the year 0 (for a set of the same projects) with respect to the mean investment I in the year 0 (for a set of the same projects), and the cash flow in the year n minus a mean cash flow (for a set of the same projects). This contribution contains discounting coefficient in the n-th order. The last term contains the contribution of the cash flow risk (as standard deviation) in the n-th year and the cash flow risk (as standard deviation) in the n'-th year and the correlation coefficient $ \rho_{n´,n} $ between the returns of the cash flow in the year n' (for a set of the same projects) with respect to the mean cash flow in the year n' (for a set of the same projects), and the cash flow in the year n minus a mean cash flow (for a set of the same projects) for the same year.

We can distinguish several cases in the aim function (\ref{0.9}) to understand better the processes behind it. For simplicity we can consider the case in which $ \rho_{I,n} = 0 $, i.e. the returns of investment in the year 0 and the returns of the cash flow in the year n are uncorrelated. This case is not realistic too much, however using it we can discuss properties of the aim function F of investors investing into projects easier, and in some cases this assumption is not so much unrealistic: when the cash flow fluctuates too much for the year n for different firms with the same project. This may happen in the industry in which there are firms with large dispersion of cash flows in the year n around some mean value near zero. In general however this mean value may be expected to be positive. This case will be discussed later on.

For simplicity we can consider also the case in which there is almost no correlation between cash flow returns in the year n and in the year n' for general n and n', $ \rho_{n´,n} = 0 $, . The curve of cash flow dependence on the year of the project may be very fluctuating around zero, and then this our assumption is realistic. In general the curve of cash flow dependence on the year of the project may be fluctuating around some value which is non-zero, mostly positive. This case we will consider latter on.

If we consider the mentioned simple cases of $ \rho_{I,n} = 0 $ and $ \rho_{n´,n} = 0 $, then the aim function (\ref{0.9}) takes a simple form:

\begin{equation}
\label{0.10}
F =  \sum_{n=1}^{n=N} \frac{{1}}{(1+r)^{n}} ( - A. CF_{n} + \frac{{\sigma_{n}}^{2}}{(1+r)^{n}} ) + ( A.I + \sigma^{2}_{I})
 \end{equation}

As we can see from (\ref{0.10}) the larger the constant A the more preferred positive cash flow $ CF_{n} $ and the larger acceptable risk $ \frac{{\sigma_{n}}^{2}}{(1+r)^{n}} $, which is however discounted. The constant A is the same for every year so for the year 0 the larger the constant A the smaller preferred investment I and the smaller risk of the investment I, assuming positive investment I.

\section{Summary and Discussion}

An investor is estimating NPV (net present value) of a firm project. Then he/she usually performs risk analysis. In \cite{SAP} it is described for example how software SAP for Banking controls risk on the market. It is possible to create portfolio hierarchies and make comparison based on these hierarchies. Then one can use these analyses to find that portfolio which corresponds to the particular needs of individual groups within the firm (bank). 
The software enable to create various scenarios of development and to choose that product which corresponds to the needs of the group. This approach is typical approach for estimating NPV of a firm project and taking into account risk.
There are methods which try to improve this approach. One of them was published recently \cite{NPVAR} and is using NPV-at-Risk method in infrastructure project investment evaluation. Distributions of Net Present Value (NPV) of the same projects in the same industry has usually normal distibution \cite{ND1} - \cite{ND21} in some cases, as it is observed from data. There exists evidence that this distribution may be in some cases different from the normal distribution. Thus there is an interesting problem: how the distribution of NPV of a firm project in a given industry may be characterized for a given firm project in a given industry. Another interesting problem is then how to describe NPV of a firm project taking into account the risk of the project using characteristics of the distribution of NPV of firm projects.
With the first problem we do not deal in this paper because there is not enough data for various firm projects in various industries to characterize distributions of NPV. The second problem may be solved for those firm project NPV of which is normally distributed. In this paper we assumed that the distribution of NPV of a firm project in a given industry is normal. 
This assumption is not true for all possible projects. We have used in our paper an opposite type of analysis to those described above and described in literature. We do not develop various scenarios and then choose the most convenient scenario for the investor, we will assume that the group (firm) is characterized by a risk aversion characteristic, which is intrinsic to this group (firm) and then using an aim function we may perform the analysis of the project in which the NPV, its risk and an aversion of the investor are included. This type of analysis of NPV is used in fact in analysis of NPV for investment into shares and bonds on capital markets. Note that in general an investment on a capital market may be considered also  as a firm project. It is known that on liquid developed markets in equilibrium there exists equivalence between investing into the shares financing a given firm project and investing directly into the firm project. 
In this paper we thus formulated such a type of analysis of NPV and risk of investment into project which is based on the aim 
function as in the case of shares in the Markowitz model and which takes into account risk aversion of the firm. The analysis is based on the fact that normal distribution of NPV is observed for some projects in some industries.
While this formulation is new approach how to analyse NPV and it is quite straightforward, the aim function for agents investing into firm projects /and for agents buying products of this project and reselling it further which is a kind of arbitrage process on the market with a given project in a given industry/ may be futher generalized which will not be done in this paper to describe analysis of NPV for which normal distribution is absent. 

We assumed that, for simplicity, there exists an investment I /which is a negative value if it is an investment of the investor for which other sources of investment like grants, subsidiarisation, etc. are less then his own sources, and it is positive if it is inverse situation/ in time zero, and that there is a cash flow $CF_{n} $ from the project every n-th year. The project life-time is assumed to be N years. The NPV of such an investment is calculated using well known expression.
The estimation of NPV is then such that it is assumed that cash flow $ CF_{n} $ in the year n has certainty to realize in the future with probability 1. It is assumed that the investment I has certainty to realize in the near future /time zero/ with probability 1. Investors then have to perform Risk and Sensitivity Analysis of the type described above, or similar with various scenarios. They are assuming for every scenario certainty to realize the investment and the cash flows with probability 1. Then investors consider various scenarios of the investment and cash flows for the same project. Note that we do not consider in this paper the case with investment and cash flow from different possible excluding mutually projects.
To consider which project is for a given investor the most profitable at a given expected risk, or which project is the less risky for an expected profit one should consider uncertainty in looking for such an optimal project. Uncertainty in cash flows and /in principle also in the investment in the time zero/ lead investors in fact to work with expected cash flows for  the future periods, and with expected investment I in zero time. Expectations of investors should include not only expected cash flow for every year n and for the expected investment I in the zero-th year, but also expected risk of net present value $ \sigma^{2}_{NPV} $.
 As we mentioned above most of the projects has normal distribution of NPV.  So we assumed in our paper that this distribution of NPV for projects /in a given industry/ is a normal distribution. NPV of a project contains for every year cash flow which is usually positive /however it may be also negative/ and the initial investment is a negative usually /however, as we noted it may be also a positive quantity/.

There exists a mean value of the NPV for projects of the same type in a given industry. In our paper we consider projects of the same type in one industry only.
We assume, that there is a liquid market with the project in the industry. We have found the mean value of the NPV, $ \overline{NPV} $ of the project. Then we discussed the risk of the NPV of the project.
The risk of the NPV of the project is for normal distribution of NPV a variance of the NPV of the project. We calculated the variance of the project. We consider an industry in which there are investments in the project of the same type over longer time periods. 
We considered $ NPV_{i} $,  the NPV of the project starting in time i. We assumed the liquid market with projects, so we can assume that at every time i there is a possibility to invest in the project. We averaged over projects which start in different time i. The rare case that the same project in the same industry starts exactly at the same time i was not considered here.

We have found the difference of the NPV and the mean value of NPV $ \overline{NPV} $. The expected risk of the project was then found. It is given by variance, in which there is $ \sigma_{n} $ the standard deviation of the year n cash flow, $ \sigma_{I} $ the standard deviation of the investment I in the time zero, $ \rho_{I,n} $ the correlation coefficient of the year n cash flow deviation from the average and of the investment I at time zero deviation from the mean investment at time zero, $ \rho_{n,n´} $ the correlation coefficient of the year $ n $ cash flow deviation from the average and of the year $ n´ $ cash flow deviation from the average. The risk of the cash flow and the zero time investment
was found to be a sum  of the discounted risk of the year n cash flow, of the zero time investment risk, of the risk associated with the correlation of the zero time investment I and the year $ n $ cash flow and  of the risk associated with the correlation of the year $ n´ $ cash flow and the year $ n $ cash flow. The later two terms are discounted correspondingly. Thus the risk of the NPV of the project consists of discounted contributions of the cash flow risks and correlations. The larger the rate of return in discounting coefficient, the smaller contribution from the risk of the year n cash flow, from the year n cash flow deviation from the average correlation with the zero time investment I deviation from the average and from the year n cash flow deviation from the average correlation with from the year $ n´$ cash flow deviation. The smaller the rate of return in discounting coefficient, the larger contribution from the risk of the year n cash flow, from the year n cash flow deviation from the average correlation with the zero time investment I deviation from the average and
from the year n cash flow deviation from the average correlation with from the year $ n´$ cash flow deviation.

The aim function of the investor into the project was found.
The investor expects an NPV of the project. To estimate this expected value we may use the historical NPV of the project in the industry. To describe the expected risk of the project we may use the historical value of the risk of the project. We assume thus that expected NPV and expected risk of the project do not differ too much from
recent historical data on projects with N years of their life. This assumption is good for projects for which N is larger than let us say two business cycles. This assumption is good also for projects in which N is two-three months. The assumption is expected not to be good for projects in which N is nearly one to three years because the business cycle leads to different behaviour of such projects in recent period with respect to the period in the next one to three years.

The investor is characterized by the constant A, which is nonnegative. The larger constant A the larger preference is given to the project NPV and the larger acceptable risk of the project.
The smaller constant A the smaller preference is given to the project NPV and the smaller acceptable risk of the project.

The aim function explicit form was found. In this form we have found
that there are contributions in which we have the aim-function-like contribution to the aim function, which is discounted and in which the risk of the n-th year risk is discounted in the second order. Further there is aim-function-like contribution to the aim-function which comes from the initial investment I and its risk $ \sigma^{2}_{I} $. The investment is a negative cash flow (expected) usually. The contribution of the cash flow risk (as standard deviation) in the n-th year and the investment risk (as standard deviation) and the correlation coefficient $ \rho_{I,n} $ between the returns of the investments I in the year 0 (for a set of the same projects) with respect to the mean investment I in the year 0 (for a set of the same projects), and the cash flow in the year n minus a mean cash flow (for a set of the same projects) are also present in the aim function. This contribution contains discounting coefficient in the n-th order. The last term contains the contribution of the cash flow risk (as standard deviation) in the n-th year and the cash flow risk (as standard deviation) in the n'-th year and the correlation coefficient $ \rho_{n´,n} $ between the returns of the cash flow in the year n' (for a set of the same projects) with respect to the mean cash flow in the year n' (for a set of the same projects), and the cash flow in the year n minus a mean cash flow (for a set of the same projects) for the same year, again discounted.

To understand better the aim function and the processes behind it we may consider several simplified cases. For simplicity we can consider the case in which $ \rho_{I,n} = 0 $, i.e. the returns of investment in the year 0 and the returns of the cash flow in the year n are uncorrelated. This case is not realistic too much, however using it properties of the aim function F of investors investing into projects may be understood easier. In some cases this assumption is not so much unrealistic: when the cash flow fluctuates too much for the year n for different firms with the same project. This may happen in the industry in which there are firms with large dispersion of cash flows in the year n around some mean value near zero. In general however this mean value may be expected to be positive. 

For simplicity we can consider also the case in which there is almost no correlation between cash flow returns in the year n and in the year n' for general n and n', $ \rho_{n´,n} = 0 $, . The curve of cash flow dependence on the year of the project may be very fluctuating around zero, and then this our assumption is realistic. In general the curve of cash flow dependence on the year of the project may be fluctuating around some value which is non-zero, mostly positive. 

If we consider the mentioned simple cases of $ \rho_{I,n} = 0 $ and $ \rho_{n´,n} = 0 $, then from the aim function  it follows that the larger the constant A the more preferred positive cash flow $ CF_{n} $ and the larger acceptable risk $ \frac{{\sigma_{n}}^{2}}{(1+r)^{n}} $, which is however discounted. The constant A is the same for every year. Note that for the year 0 the larger the constant A the smaller preferred investment I and the smaller risk of the investment I, assuming positive investment I. Thus we have described the aim function of NPV for investors investing in a project from an industry in the case in which the distribution of NPV is normal one. This aim function may be used in analysis of NPV and risk of a project for investors differing in preference with respect to risk. That investmet and cash flows are the most acceptable for an investor with a given constant A which give the minimum of the aim function.

\section*{Acknowledgment}
The first author has done this work during her scientific stay in Matematical - Economical Consultancy firm.
The second author was supported by the VEGA project 1/0495/03.


\begin{thebibliography}{AAA}
\bibitem{1}
T. Plummer, The Psychology of Technical Analysis, Rev. Ed., Probus Pub.Comp., Chicago-Cambridge, 1993
\bibitem{2}
D. Lewis, The Secret Language of Success, Carroll and Graf Pub. Inc., USA, 1989
\bibitem{3}
N. Rivier, Journal de Physique, C9 N12 T46 (1985) 155
\bibitem{4}
N. Rivier, Physica, 23D (1986) 129
\bibitem{5}
O. Hudak, Topology and Social Behaviour of Agents,  http://arXiv.org/abs/cond-mat/0312723 , 2003
\bibitem{6}
I. Molho, The Economics of Information, Blackwell Pub., Oxford, Malden, 1997
\bibitem{SAP} see on http://www.sap.com
\bibitem{NPVAR}S. Ye and R.L.K. Tiong, NPV-at-Risk Method in Infrastructure Project Investment Evaluation, Journal of Construction Engineering and Management, Vol. 126, No. 3, May/June 2000, pp. 227-233 
\bibitem{ND1}Ch.-F. Lee, Statistics for Business and Financial economics, World Scientific, 1998, p. 273
\bibitem{ND2}J. Sabal, Financial Decisions in Emerging Markets, Oxford Univ. Press, 2002, p. 42
\bibitem{ND3}J.L. Grant, J.A. Abate, Focus on Value: A Corporate and Investor Guide to Wealth Creation, J. Wiley and Sons, 1996, p. 117
\bibitem{ND4}H.L. Beenhakker, Investment Decision Making in the Private and Public Sectors, Greenwood Pub. Group, 1996, p. 159
\bibitem{ND5}P. Fernandez, Valuation Methods and Shareholder Value Creation, Elsevier, 2002, p. 551
\bibitem{ND6}H. Khatib, Economic Evaluation of Projects in the Electricity Supply Industy, IEE Power and Energy Series 44, 2003, p. 180
\bibitem{ND7}K. Wang, M.L. Wolverton, Real Estate Valuation Theory, Springer, Research Issues in Real Estate Vol. 8, 2002, p. 419
\bibitem{ND8}S. Harrison, J. Herbohn, R. Irons, P. Rowland, Capital Budgeting> Financial Appraisal of Investment Projects, Cambridge Univ. Press, 2003, p. 177
\bibitem{ND9}J.G. Siegel, J.K. Shim, Finance, Barons Educational Series EZ-102, 1991, p. 65
\bibitem{ND10}H.L. Beehakker, The Global Economy and International Financing, Greenwood Pub. Group, 2000, p. 145
\bibitem{ND11}J. Fisher, P. Mosbaugh, Language of Real estate appraisal, Dearborn, Real Estate Education, 1990, p. 129
\bibitem{ND12}W.K. Brauers, Optimization Methods for a Stakeholder Society: A Revolution in Economic Thinking by Multi-Objects, Springer, 2004, p. 92
\bibitem{ND13}M. Loosemore, T.E. Uher, Essentials of Construction Project management, UNSW Press, 2003, p. 352
\bibitem{ND14}L.H. Jacobs, The Revolution in Corporate Finance, Blackwell Pub., 2003, p. 92
\bibitem{ND15}J.N. Luftman, Competing in the Information Age: Align in the Sand, Oxford Univ. Press, 2002, p. 103
\bibitem{ND16}R.Ch. Moyer, J.R. McGuigan, W.J. Kretlow, ContemporaryFinancial management, Thomson South-Western, 2005, p. 379
\bibitem{ND17}Ch. Narrison, The Fundamentals of Risk Measurements, McGraw Hill Professional, 2002, p. 394
\bibitem{ND18}P. Belli, J.R. Anderson, H.N. Barnum, J.A. Dixon, J.-P. Tan, Economic Analysis of Investment Operations:
Analytical Tools and practical Applications, The World Bank, 2000, p. 156
\bibitem{ND19}C.S. Patterson, The Cost of Capital: Theory and Estimation, Greenwood Pub. Group, 1995, p. 296
\bibitem{ND20}R. Brown, H. Campbell, Benefit-Cost Analysis: Financial and Economic Appraisal using Spreadsheets, Cambridge Univ. Press, 2003, p. 201
\bibitem{ND21}J.C. Goodpasture, Quantitative Methods in Project Management, J. Ross Pub., 2003, p. 263
\bibitem{7}
A. Greenspan, Commercial Banks and the Central Bank in a Market Economy, in Readings on Financial Institutions and Markets, P.S. Rose editor, 5th ed., R.D. Irwin Inc., Homewood, Boston, 1993, p. 294
\bibitem{8}
J. von Neumann and O. Morgenstern, Theory of Games and Economic Behaviour, 3rd. ed., Princeton University Press, Princeton, 1953
\bibitem{9}
J. Lintner, The Market Price of Risk, Size of Market and Investor´s Risk Aversion, Journal of Business, April (1968)
\bibitem{10}
W.F. Sharpe, Integrated Aset Allocation, Financial Analysts Journal, September - October (1987)
\end{thebibliography}
\end{document}